\documentclass[iop]{emulateapj}
\usepackage{graphicx}
\usepackage{subfigure,aas_macros,comment}

\newcommand\msun{\rm{M_{\odot}}}

\def\stacksymbols #1#2#3#4{\def\theguybelow{#2}
        \def\verticalposition{\lower#3pt}
        \def\spacingwithinsymbol{\baselineskip0pt\lineskip#4pt}
        \mathrel{\mathpalette\intermediary#1}}
\def\intermediary #1#2{\verticalposition\vbox{\spacingwithinsymbol
        \everycr={}\tabskip0pt
        \halign{$\mathsurround0pt#1\hfil##\hfil$\crcr#2\crcr
                \theguybelow\crcr}}}

\def\lta{\stacksymbols{<}{\sim}{2.5}{.2}}
\def\gta{\stacksymbols{>}{\sim}{2.5}{.2}}

\shortauthors{M. Gaspari et al.}
\shorttitle{Breaking galaxies, groups, and clusters with AGN feedback}

\begin{document}

\title{Can AGN feedback break the self-similarity of galaxies, groups, and clusters?}
\author{M. Gaspari$^{1,2,5}$, F. Brighenti$^{2,3}$, P. Temi$^4$, S. Ettori$^{5,6}$}
\affil{$^{1}$Max Planck Institute for Astrophysics, Karl-Schwarzschild-Strasse 1, 85741 Garching, Germany; mgaspari@mpa-garching.mpg.de\\
$^{2}$Astronomy Department, University of Bologna, Via Ranzani 1, 40127 Bologna, Italy\\
$^{3}$UCO/Lick Observatory, Department of Astronomy and Astrophysics, University of California, Santa Cruz, CA 95064, USA\\
$^{4}$Astrophysics Branch, NASA/Ames Research Center, MS 245-6, Moffett Field, CA 94035\\
$^{5}$INAF, Osservatorio Astronomico di Bologna, via Ranzani 1, 40127 Bologna, Italy\\
$^{6}$INFN, Sezione di Bologna, viale Berti Pichat 6/2, 40127 Bologna, Italy}

\begin{abstract}
\noindent
It is commonly thought that AGN feedback can break the self-similar scaling relations of galaxies, groups, and clusters. 
Using high-resolution 3D hydrodynamic simulations, we isolate the impact of AGN feedback on the $L_{\rm x}-T_{\rm x} $ relation, testing the two archetypal and common regimes, self-regulated mechanical feedback and a quasar thermal blast. We find that AGN feedback has severe difficulty in breaking the relation in a consistent way.
The similarity breaking is directly linked to the gas evacuation within $R_{500}$, while the central cooling times are inversely proportional to the core density.
{\it Breaking self-similarity implies thus breaking the cool core}, morphing all systems to non-cool-core objects, which is in clear contradiction with the observed data populated by several cool-core systems.
Self-regulated feedback, which quenches cooling flows and preserves cool cores, prevents the dramatic evacuation and  similarity breaking at any scale; the relation scatter is also limited. The impulsive thermal blast can break the core-included $L_{\rm x}-T_{\rm x}$ at $T_{500}\lta 1$ keV, but substantially empties and overheats the halo, generating a perennial non-cool-core group, as experienced by cosmological simulations. Even with partial evacuation, massive systems remain overheated. We show the action of purely AGN feedback is to lower the luminosity and heating the gas, perpendicular to the fit. 
\end{abstract}


\section{Introduction}\label{s:intro}
\noindent
In the last decade, feedback due to active galactic nuclei (AGN) has allowed to solve crucial astrophysical problems.
The supermassive black hole (SMBH) at the center of galaxies, groups, and clusters
can indeed release a terrific amount of energy ($>10^{61}$ erg), providing an efficient source to quench cooling flows and star formation (\citealt{McNamara:2007}). 
In particular, mechanical AGN feedback in the form of jets/outflows is able to regulate for several Gyr the thermodynamical state of the system core (\citealt{Gaspari:2013_rev} for a review).
However, it is far from clear if AGN feedback is able to strongly modify the large-scale gas halo, as the total X-ray luminosity and temperature, in other words, breaking the self-similar scaling relations.

If gravity were the single driver of the evolution (\citealt{Kaiser:1986, Kravtsov:2012}), all systems would scale only with mass, $M_{\Delta} = (4\pi/3) \Delta\,\rho_{\rm c}\,R^3_\Delta$, where $\Delta$ is the chosen overdensity. 
The critical density of the universe evolves in redshift as $\rho_{\rm c}(z)\propto E^2(z)$, where $E^2(z)\simeq\Omega_{\rm m}\,(1+z)^3 +\Omega_\Lambda$, giving a characteristic radius $R_{\Delta}\propto M_{\Delta}^{1/3}\, E^{-2/3}(z)$. Via hydrostatic equilibrium ($M_\Delta\propto T\, R_\Delta$), we can retrieve $M_{\Delta}\propto T^{3/2}E^{-1}(z)$. Since the bolometric X-ray luminosity scales as $L_{\rm x} \propto n^2\, T_{\rm x}^{1/2}\, R^3_{\Delta}$ (in the Bremsstrahlung regime), using gas number density $n\propto M_{\Delta}/R_{\Delta}^3\propto\rho_{\rm c}(z)$ and the above relations, we find the well-known self-similar scaling $L_{\rm x}\propto T_{\rm x}^2\,E(z)$.
However, cluster observations show a slope steeper than 2 ($\sim$3; e.g., \citealt{Pratt:2009,Maughan:2012}), further sharpening in the group regime, $\sim$4$\,$-$\,$5 (\citealt{Mulchaey:2003, Osmond:2004, Helsdon:2000b, Helsdon:2000a, Sun:2009a,Sun:2012}; see Fig. \ref{fig1}).

In recent years, different authors have studied the scaling relations by means of large cosmological simulations with AGN feedback (e.g., \citealt{Sijacki:2007,Fabjan:2010,McCarthy:2010,Short:2010}). 
In general, they find that the implemented AGN feedback is able to break the self-similarity, lowering luminosities by orders of magnitude and, surprisingly, decreasing the global temperature (\citealt{Puchwein:2008}, fig.~2). 
Often overlooked, 
the simulated systems are however non-cool-core objects (\citealt{Planelles:2013}, fig.~7, for a critical discussion).
Even no-feedback runs produce negative temperature gradients, due to extreme adiabatic heating,
which are not present in high-resolution simulations (e.g., \citealt{Li:2012}). 
Besides the under-resolved black hole/feedback physics and subgrid numerics
(see the analysis in \citealt{Barai:2014}),
it remains difficult to disentangle and isolate the action of feedback in the complex evolution shaped by mergers, filaments,
star formation, sink particles, and other prescriptions.

The objective of this study is to critically examine if AGN feedback itself can break the self-similarity of galaxies, groups, and clusters, in a way consistent with observations. Via controlled high-resolution 3D (mesh) simulations, we study how the main scaling, $L_{\rm x}-T_{\rm x}$ is shaped by the two archetypal and commonly adopted feedback models, self-regulated kinetic feedback and a quasar thermal blast. In a forthcoming work, we explore other models and different relations (Gaspari et al. 2014, in prep.).

A crucial constraint driven by observations is the presence of a (strong or weak) cool core in the majority of observed
systems ($\gta65$ per cent, \citealt{Peres:1998, Mittal:2009, Sun:2009a,Hudson:2010,Zhao:2013}).
Such systems show, in the core, cooling times $< 7$ Gyr, positive temperature gradients, and low gas entropy ($<\,$10s keV cm$^2$). Moreover, cool cores appear to be long lived and in place since $z > 1$ (\citealt{McDonald:2013}).
AGN feedback, or inside-out heating, intrinsically evacuates the central regions, before touching the periphery of the system. Although the AGN feedback energetics is in principle capable to breaking the group self-similarity (\citealt{Cavaliere:2008,Giodini:2010}), the energy deposition and hydrodynamics is crucial.
We show that breaking self-similarity via AGN feedback implies disrupting the cool core, morphing the system into perennial non-cool-core objects; vice versa, self-regulation preserves the core and the large-scale structure.

\section[]{Physics \& Numerics} \label{s:num}

\subsection[]{Initial conditions} \label{s:init}
\noindent
In order to fully isolate the role of feedback in altering the scaling relations, we start with a virialized group/cluster having a formed cool core, which characterizes the majority of observed systems (\S\ref{s:intro}). 
Groups and clusters share many common properties, allowing to build an 
initial `universal' system defined only by its mass.
Following \citet{Vikhlinin:2006}\footnote{Note a typo in the published version, missing the 0.45 exponent; A.~Vikhlinin, private communication (see astro-ph version).},
the observed average temperature profile can be modeled as
\begin{equation}\label{Tfit}
T(r) = T_0\; \frac{0.45+(\hat{r}/0.045)^{1.9}}{1+(\hat{r}/0.045)^{1.9}} \,\frac{1}{(1+(\hat{r}/0.6)^2)^{0.45}}, 
\end{equation}
where $\hat{r}=r/R_{\rm 500}$; the normalization is $T_0 \simeq 1.4\; T_{500}
\simeq 3\ {\rm keV}\, (M_{500}/10^{14}\, \msun)^{0.6}$ (cf., \citealt{Sun:2009a}). Eq.~\ref{Tfit} models the positive gradient of the cool core and the gentle decrease at large $r$; the peak temperature ($r\sim0.15\,R_{500}$) is $\sim$2$\times$ the central value, which is reached again at $r\sim R_{500}$.
Albeit some groups have slightly higher $T$ peak and steeper decrease (\citealt{Sun:2009a}),
such minor differences have no impact on the results. 

The system is initially in hydrostatic equilibrium within the gravitational potential $\phi$, dominated by dark matter,
modeled via the usual NFW profile in the concordance $\Lambda$CDM universe.
The halo concentration is linked to the virial mass as $c \simeq 8.5\, (M_{\rm vir}/10^{14}\ \msun)^{-0.1}$ (e.g., \citealt{Bullock:2001}).
In addition, each group/cluster is dominated by a central massive elliptical galaxy (`BCG'), modeled 
with a de VaucouleursÕ stellar density profile.
The BCG $K$-band luminosity increases with the halo mass as $L_K\simeq4.7\times10^{11}\, (M_{\rm vir}/10^{14}\ \msun)^{0.39}\,L_\odot$ (e.g., \citealt{Lin:2004}). The stellar mass is then retrieved adopting $M_\ast/L_K\sim1$ (e.g., \citealt{Mannucci:2005}). 
Since BCGs are large ellipticals, we keep the effective radius $R_{\rm eff}\simeq 9$ kpc. 
As described in \citet{Gaspari:2012b},
the BCG injects a low amount of energy and mass due to SNIa and stellar winds; however,  
the energetics is dominated by the AGN feedback.

The normalization of the density profile is set by the gas fraction at the virial radius, $f_{\rm gas, vir} \simeq 0.15$. 
The initial gas fraction is intentionally high, near the cosmic value (\citealt{Planck:2013_params}), since we want to test if AGN feedback is the original cause of gas evacuation and hence self-similarity breaking. 
Using lower values (e.g., 0.1) does not change the conclusions.

\subsection[]{Hydrodynamics, cooling and heating}\label{s:hydro}
\noindent
Using FLASH4 code, we integrate the 3D equations of hydrodynamics in conservative form, including total gravity, gas radiative cooling, and feedback heating. The latter two source terms are implemented 
following the unified self-regulation model, as described in \citet{Gaspari:2012b}.
Transport mechanisms, as conduction, are not included since data suggest a strong suppression (e.g., \citealt{Gaspari:2013_coma}).
The cubical box fully covers the virial radius, $\sim$1.2$\,$-$\,$4.6 Mpc (groups to clusters).
We use concentric grid levels with radius of $\sim$60 cells, centered on the BCG, where the maximum resolution reaches $\approx 290$ pc. The system is integrated for at least 5 Gyr. Boundary conditions are set in diode mode.

\subsubsection{Self-regulated mechanical feedback} \label{s:cold}
\noindent
In \citet{,Gaspari:2011a,Gaspari:2011b,Gaspari:2012a,Gaspari:2012b,Gaspari:2013_rev,Gaspari:2013_CCA} was found that the most consistent model able to solve the cooling flow problem is mechanical feedback, 
self-regulated by cold accretion. 
In turbulent regions where the cooling time drops below $\sim$$10\times$ the free-fall time,
thermal instabilities become quickly nonlinear, leading to the condensation of cold gas out of the hot phase. 
Such cold clouds and filaments 
collide in an inelastic and chaotic way while raining on to the black hole, boosting the accretion rate.  
Bipolar massive sub-relativistic outflows are then triggered with kinetic power proportional to the central cooling rate, $P_{\rm jet} = \epsilon\, \dot{M}_{\rm cool} \,c^2$ (\citealt{Gaspari:2012b} for the numerical details), with optimal mechanical efficiencies $\epsilon\sim5\times10^{-4}-5\times10^{-3}$. 
The self-regulated outflow generates the cocoon shock, two buoyant bubbles, and gas/metal uplift. The kinetic feedback rises the central gas entropy, quenching cooling and stifling the accretion rate; the self-regulated loop starts then over again.
 
The gentle self-regulation with either kinetic or thermal injection (the latter commonly used in cosmological simulations\footnote{In some cosmological works, the `quasar mode' is simply quasi-continuous thermal feedback.})
produces analogous impact on the scaling relations, although thermal feedback induces again excessive core overheating
(\citealt{Brighenti:2003, Gaspari:2011b}).

\begin{figure*} 
    \begin{center}
       \subfigure{\includegraphics[scale=0.34]{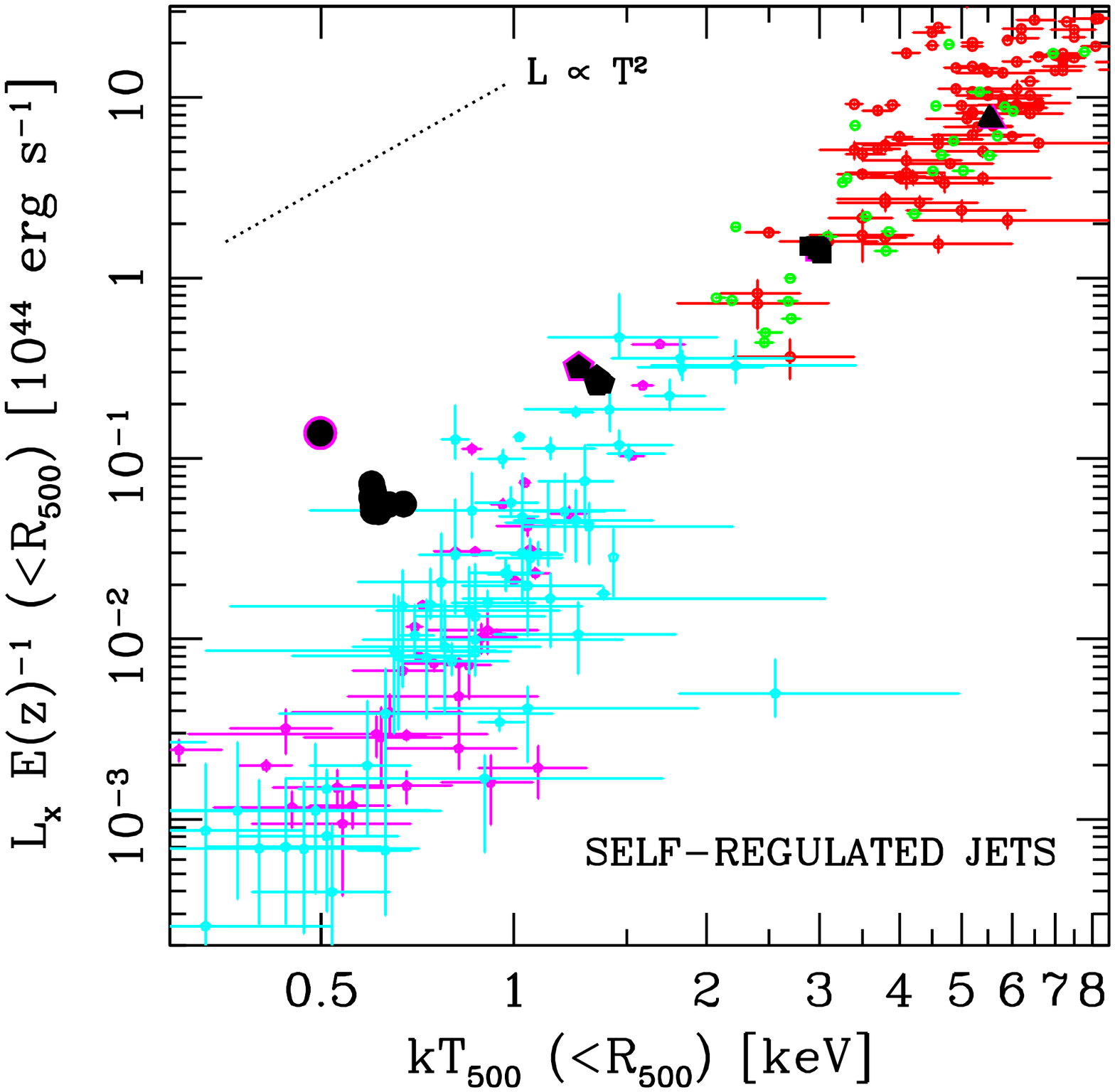}}
       \hspace{+1cm}
       \subfigure{\includegraphics[scale=0.34]{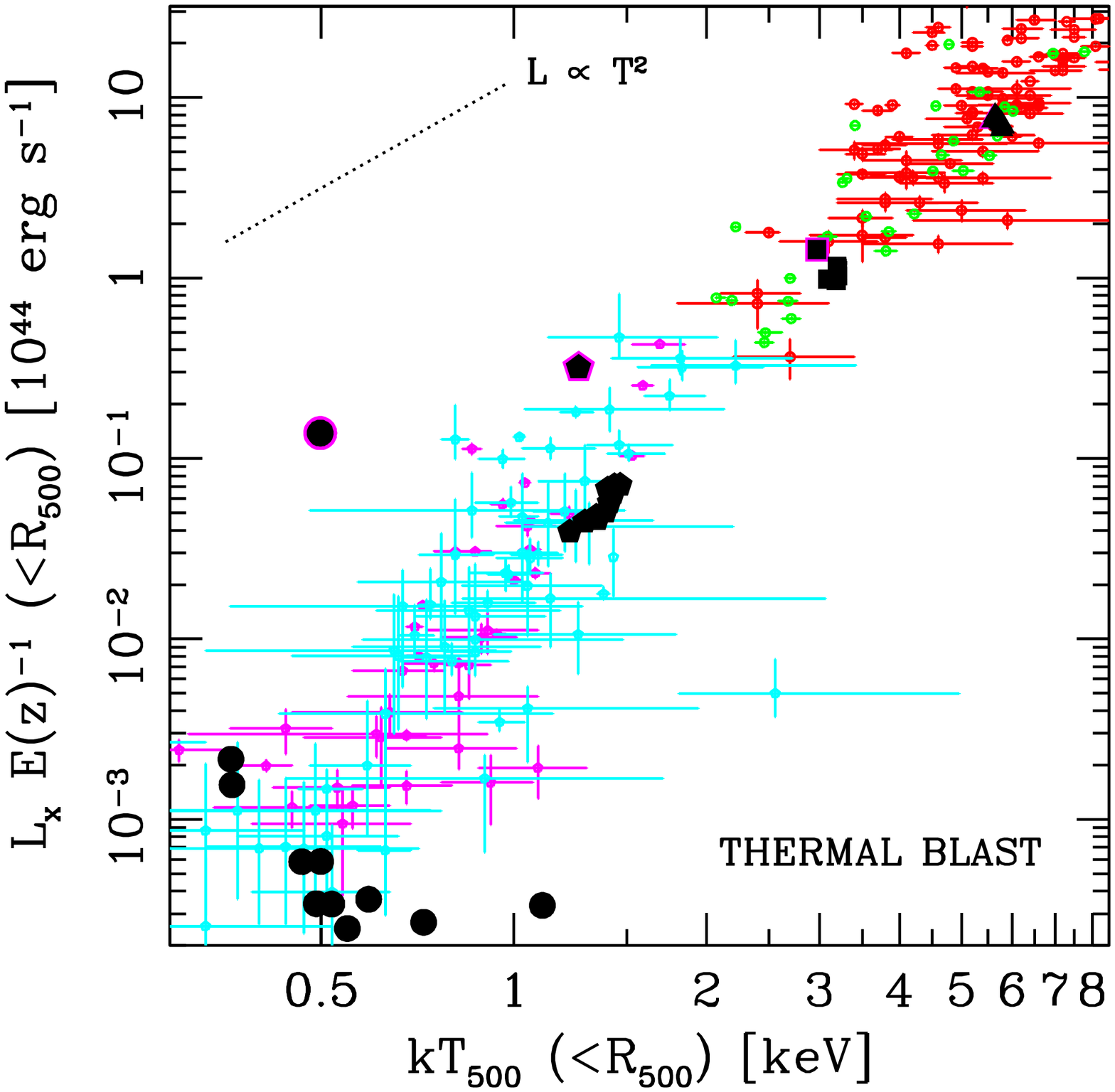}}       
       \subfigure{\includegraphics[scale=0.34]{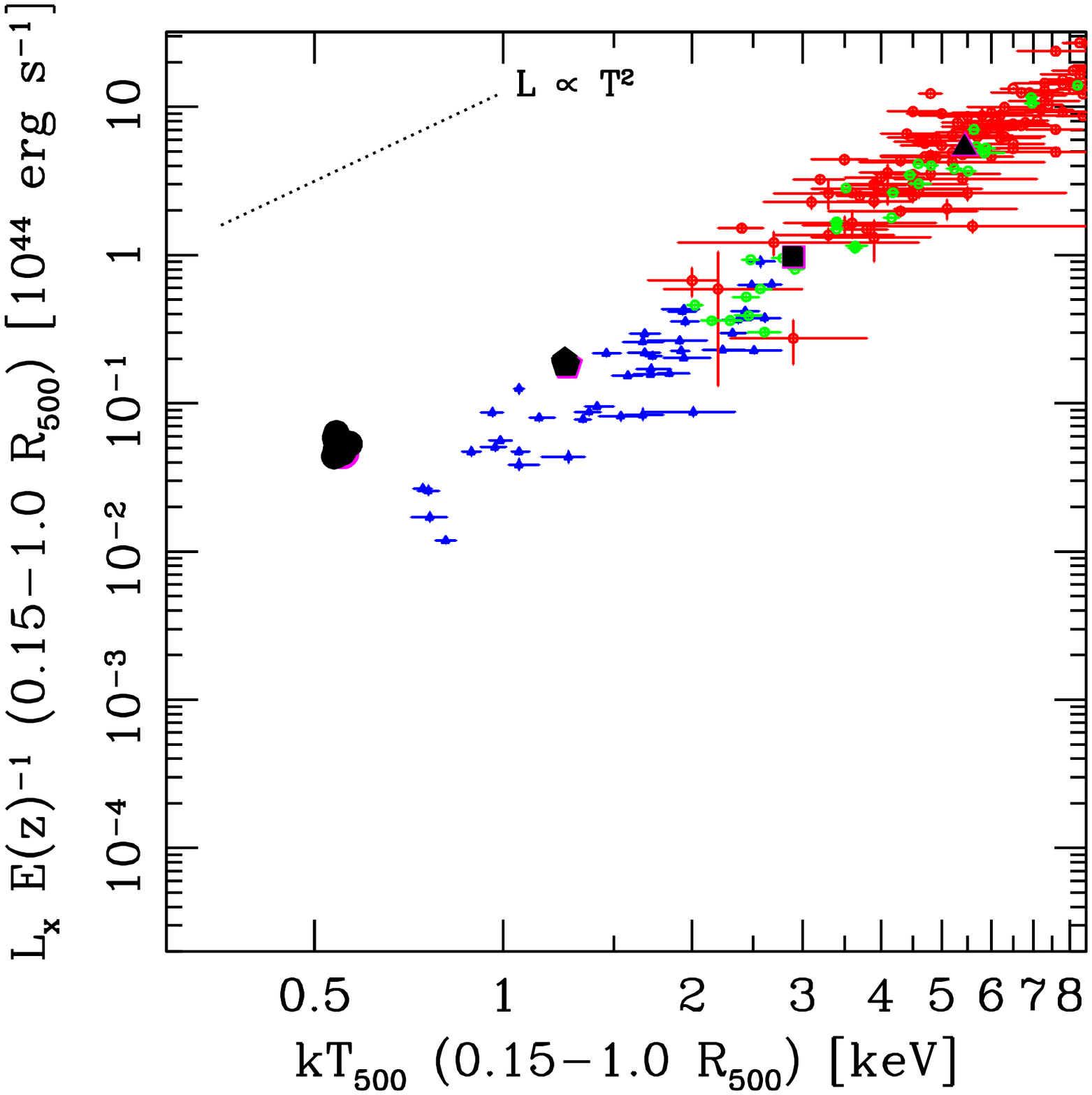}}
       \hspace{+1cm}
       \subfigure{\includegraphics[scale=0.34]{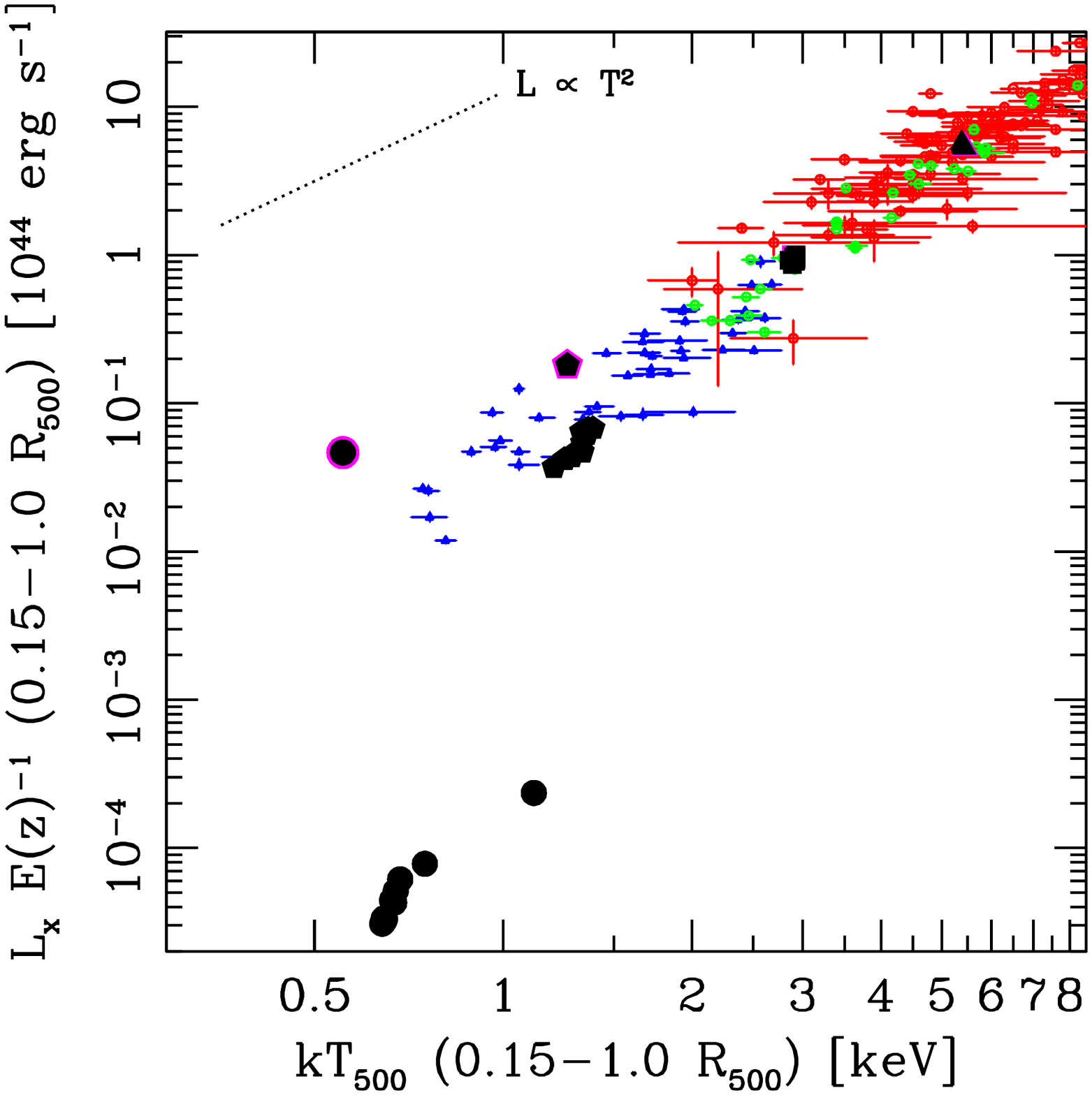}}       
       \subfigure{\includegraphics[scale=0.34]{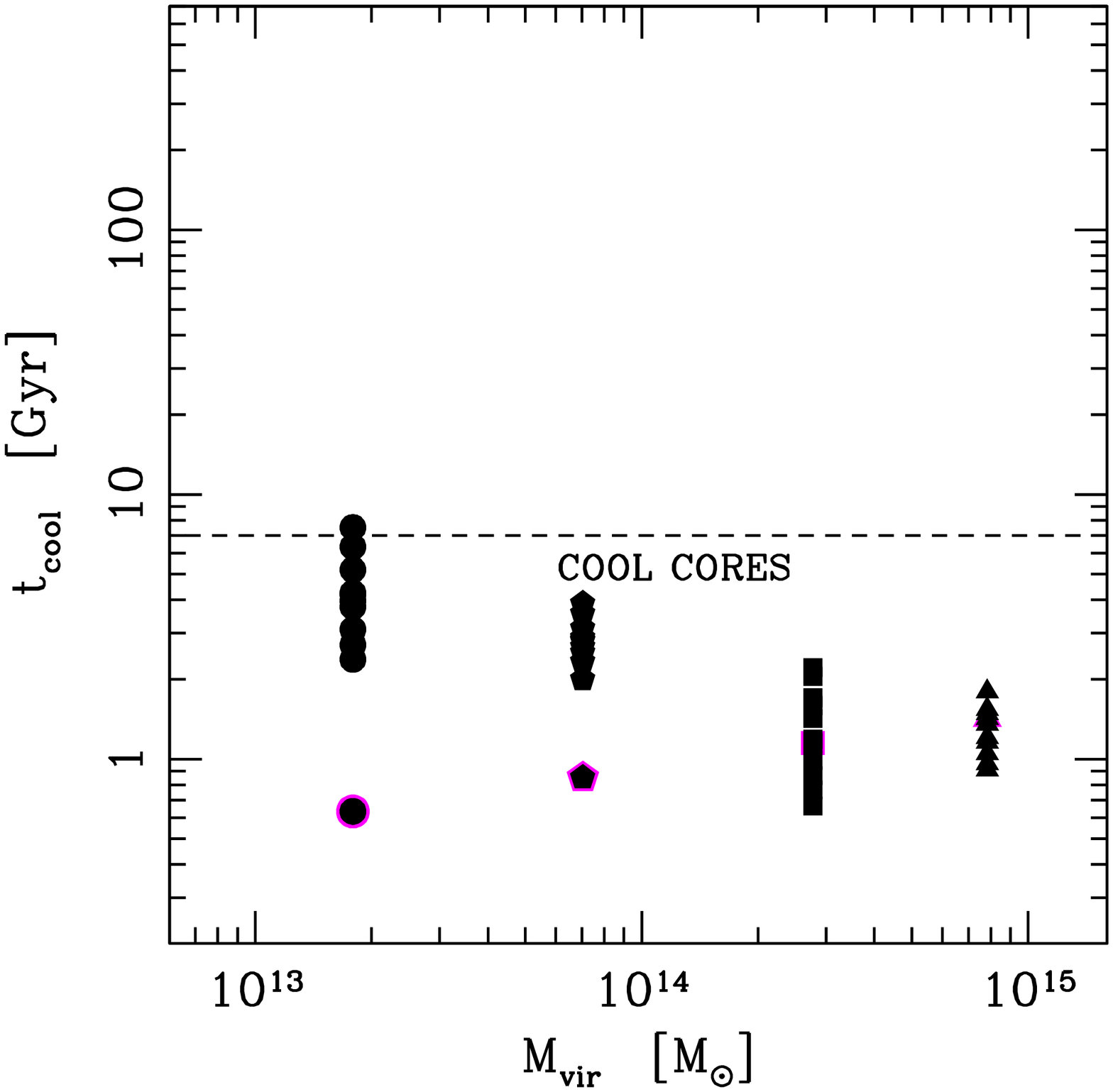}}      
       \hspace{+1cm}
       \subfigure{\includegraphics[scale=0.34]{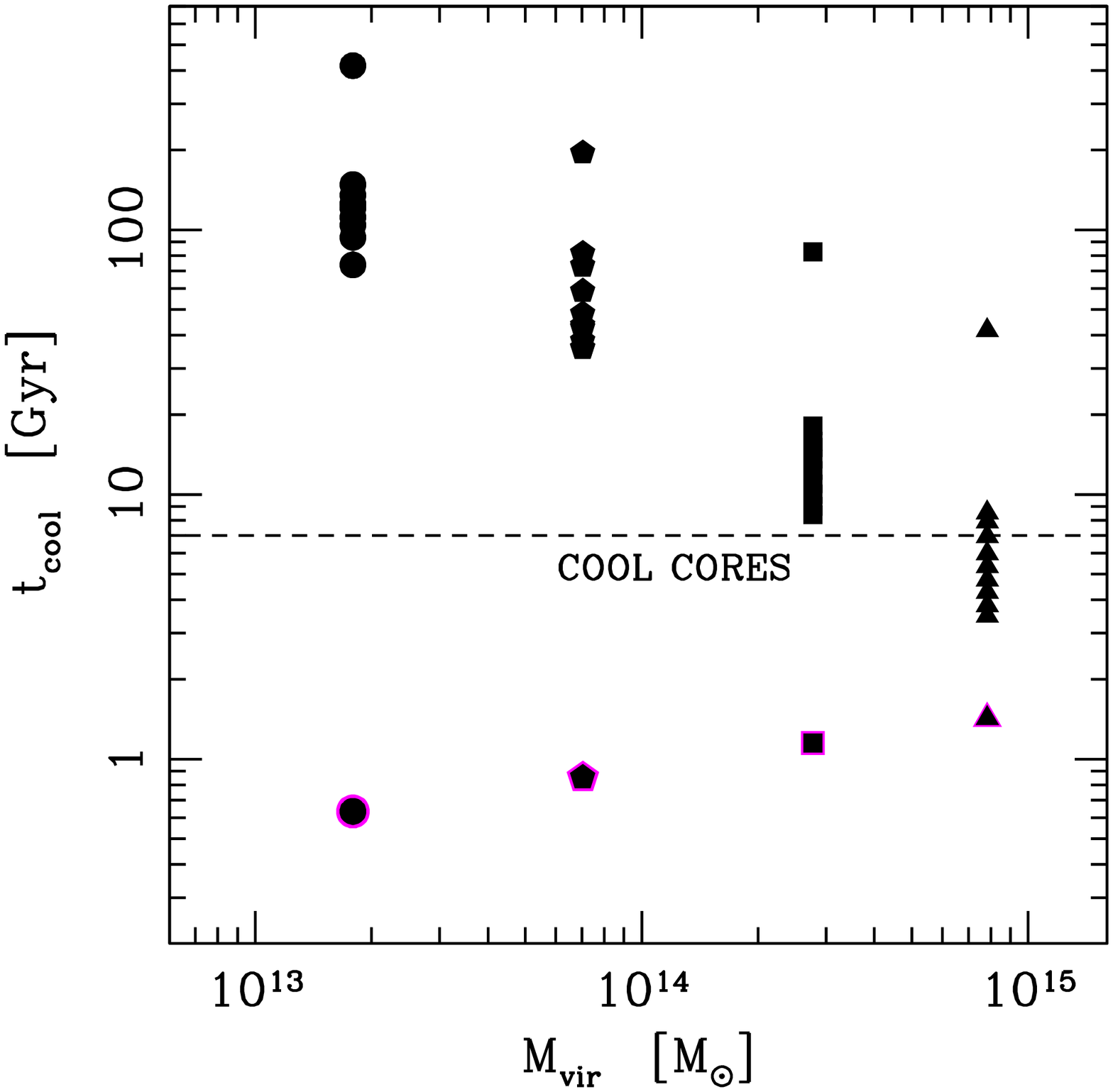}}         
     \end{center}
     \caption{X-ray bolometric luminosity versus X-ray temperature, including (top) or excising (middle) the core, $r<0.15\,R_{500}$. The bottom panels show the central cooling time ($\sim$15-20 kpc), with the cool-core threshold $t_{\rm cool}\,$$\sim\,$7 Gyr. Left: self-regulated kinetic models (\S\ref{s:cold}; $\epsilon = 5\times10^{-3}$). Right: quasar thermal blast (\S\ref{s:thermal}); the black points show the simulated 5 Gyr evolution every 500 Myr (the initial point has magenta contour).
The observational data are from \citeauthor{Maughan:2012} (2012, {\it Chandra}; red), \citeauthor{Pratt:2009} (2009, {\it XMM}; green), \citeauthor{Sun:2009a} (2009, {\it Chandra}; blue), \citet{Mulchaey:2003} and \citeauthor{Osmond:2004} (2004; magenta), \citeauthor{Helsdon:2000a} (\citeyear{Helsdon:2000b},b, {\it ROSAT}; cyan); we always use $h=0.7$ (e.g., for $L_{\rm x}\propto h^{-2}$). 
The self-regulated AGN feedback prevents overheating, at the same time avoiding the self-similarity breaking.
Conversely, the powerful thermal blast can break the $L_{\rm x}-T_{\rm x}$ relation at the group scale, but morphing the system into a perennial non-cool-core object.
          \label{fig1}}
\end{figure*}  

\subsubsection{Quasar thermal blast} \label{s:thermal}
\noindent
In the opposite spectrum of feedback models resides the sudden and powerful release of thermal energy.
This can be justified by a quasar event, emitting large radiative power absorbed by highly dense clouds, or in alternative, by an Eddington wind fully thermalizing in the inner core.
Numerically, the thermal energy is injected in the inner $\sim$4 kpc, 
with Eddington power $P_{\rm Edd}\simeq1.5\times10^{47}(M_{\rm bh}/10^9\,\msun)$ erg s$^{-1}$ lasting $\sim\,$6 Myr. The total released energy is $E_{\rm AGN}\equiv\eta \,M_{\rm bh}\,c^2\simeq3\times10^{61}$ erg, i.e., the characteristic energy of a SMBH with typical $M_{\rm bh}\sim10^{9}\ \msun$ and radiative efficiency $\eta \simeq 1.5\times10^{-2}$ (\citealt{Novak:2013}). The isotropic blast triggers once the necessary mass of cold gas has been accreted; due to the powerful heating, no second event ever occurs.
Conclusions are unaltered with different $E_{\rm AGN}$ and compact deposition windows.

Boosting $\epsilon$ above the optimal values (e.g., $\sim0.1$) transforms the previous gentle self-regulated feedback in the impulsive blast. Conversely, significantly lowering $E_{\rm AGN}$ morphs the feedback in the self-regulated regime. 
The presented models constitute thus the two opposing archetypes of inside-out feedback.

\section[]{Results} \label{s:res}
\noindent
Figure \ref{fig1} presents the key results of
the high-resolution hydrodynamic simulations, testing kinetic or thermal AGN feedback in the range of systems with 
$T_{500}\simeq0.5\,$-$\,6$ keV ($M_{\rm vir}\sim10^{13}\,$-$\,10^{15}\ \msun$).
In the top and middle panels, we show the X-ray luminosity versus X-ray temperature\footnote{
We computed both the emission-weighted $T_{500}$ with {\it Chandra} sensitivity ($T_{\rm x}\gta 0.3$ keV; \citealt{Gaspari:2012b}) and spectroscopic-like temperature (\citealt{Vikhlinin:2006_temp}); since our flow is not multiphase, they are very similar. As in observations, we use the projected $T_{500}$; the difference with the spherical value is minor.} within $R_{500}$,
including or excising the core ($r<0.15\,R_{500}$), respectively. The bottom panel depicts the gas central cooling time (in the shell $\sim$15$\,$-$\,$20 kpc, contained within $\lta0.06\, R_{500}$). The $L_{\rm x}-T_{\rm x}$ relation is shaped by the global amount of cooling and heating, while 
$t_{\rm cool}\propto T/n\,\Lambda$ assesses the core thermal state ($\Lambda$ is the cooling function; see \citealt{Gaspari:2012b}).

The self-similar relation is expected to be $L_{\rm x}\propto T_{\rm x}^2$, even shallower in the group regime due to line emission ($L_{\rm x}\propto T_{\rm x}^{3/2}\Lambda \propto T_{\rm x}$).
Figure \ref{fig1} (top) reveals that the observational data relative to the cluster regime (red, green) are already deviating, with a slope $\alpha\sim3$. Below 2 keV, i.e., for small and massive groups, the relation steepens further, reaching $\alpha\sim\,$ 4$\,$-$\,$5, with a much more significant scatter. Excising the core in both quantities reduces the scatter and the steepness ($\alpha\sim2.5$), avoiding an abrupt decline. 
The relation becomes nearly self-similar considering only cool-core clusters (\citealt{Maughan:2012}). 
Unfortunately, the excised relation for small groups is not covered by observational data.

In the left column, the simulations (black; the initial state has magenta contour) 
show that the impact of self-regulated jet feedback ($\epsilon=5\times10^{-3}$) on the $L_{\rm x}-T_{\rm x}$ relation is limited. The remarkable aspect is that no break -- a deviation of orders of magnitude -- occurs, even at the scales of small halos.
In the massive group (pentagons), the maximum deviation in luminosity/temperature is $\sim\,$30/10 per cent (0.1/0.05 dex),
strongly diminishing to $\sim$10/1 per cent in massive clusters. In the compact group (circles), the luminosity decreases by $\sim$$2.8\times$, while temperature increases by $\sim$80 per cent. $T_{500}$ shows the weaker scatter;
the main action of feedback, especially kinetic, is to evacuate gas and, secondarily, to heat the global atmosphere. 

Excising the core within $0.15\,R_{500}$ (middle panel), significantly reduces the scatter to $\lta\,$1/3 of the previous values.
By removing the core separately for each variable, we see that internal heating can only move the points towards lower luminosities and higher temperatures. Similarly, radiative cooling moves the system towards higher $L_{\rm x}$ and lower $T_{500}$ (cf., \citealt{Ettori:2008}). In other words, the $L_{\rm x}-T_{\rm x}$ secularly moves due to heating/cooling perpendicular to the fit, particularly as the core is included. 
Works based on cosmological simulations (\S\ref{s:intro}) show instead a decrease of $T_{500}$,
adding subgrid AGN feedback. This could be linked to the reduction of the extreme adiabatic heating present
in the under-resolved pure cooling flow.
A more serious problem is that, although the similarity breaking occurs, no object can be observationally described as a cool core with the positive $T$ gradient depicted in \S\ref{s:init}. 

The self-regulated models show that avoiding the complete self-similarity breaking implies preserving the cool-core structure -- at the same time reducing the cooling rate below 10 per cent of that of the pure cooling flow.
In the bottom panel (left), the central cooling time stays in any halo below $\sim$7 Gyr, the common upper limit
used to define cool-core systems (e.g., \citealt{Hudson:2010}). 
These systems also preserve the positive temperature gradient and low central entropy.
As found in \citet{Gaspari:2011b,Gaspari:2012b},
$\epsilon \simeq5\times10^{-3}$ is the best value for clusters; indeed, the two groups switch to a weaker cool core after the initial heating.
Overall, optimal self-regulation induces the system to oscillate between a state of weak and strong cool core, preventing both the cooling {\it and} heating runaway. The duty cycle is very efficient with cold accretion, while much weaker with hot Bondi regulation (\citealt{Gaspari:2011b}). 
Since most of the observed systems host a cool core (\S\ref{s:intro}), self-regulated mechanical heating represents the long-term maintenance mode of AGN feedback, while avoiding the break of the scaling relations.

We test now the other extreme of AGN feedback models, i.e., the impulsive thermal blast (Fig. \ref{fig1}, right). 
The sudden isotropic energy release ($\sim$$3\times10^{61}$ erg, comparable to that of a typical SMBH)
can dramatically evacuate the atmosphere of the compact group, decreasing the X-ray luminosity by 2.5 orders of magnitude; $T_{500}$ initially increases by $\sim\,$$2\times$ (the rightmost circle).
The strong breaking occurs because the group total binding energy\footnote{Equal to the gravitational energy $E_{\rm b}=\int_0^{R_{\rm vir}}{\rho_{\rm gas}\,\phi\; dV}$.} is $\simeq10^{61}$ erg. After the initial blast, the BCG is slowly replenished by the stellar mass loss, mildly increasing the luminosity ($<\,$$R_{500}$) to $\sim10^{41}$ erg s$^{-1}$, while restoring $T_{500}$ near the initial value (this is not due to the action of AGN feedback).
The system has completely morphed. The central $t_{\rm cool}$ (bottom) is always $>\,$60 Gyr, in a perennial non-cool-core state. Excising the core (middle), aggravates the breaking, since the compact group is substantially devoid of gas outside the BCG, with an unrealistic drop
down to $\sim\,$$10^{40}$ erg s$^{-1}$ 
(cf., \citealt{Sun:2012}). Overall, the inside-out heating able to {\it fully} break self-similarity in the core-included $L_{\rm x}-T_{\rm x}$,  violently alters the excised $L_{\rm x}-T_{\rm x}$ relation, which is instead observed to have tighter scatter.

In the massive group (pentagons), the total binding energy is $\simeq10^{62}$ erg ($> E_{\rm AGN}$). The thermal blast can thus only partially evacuate the gas from the core ($\sim0.1\, R_{500}$), halted by the extended atmosphere before reaching $R_{500}$.
The result is a decrease in luminosity by maximum 0.9 dex and $T_{500}$ oscillating within 0.1 dex. The extended evacuation is confirmed by the excised relation.
Both simulated $L_{\rm x}-T_{\rm x}$ are consistent with the observed data. However, the similarity deviation occurs again at the expense of the cool core. In fact, the central cooling time stays above tens Gyr, signaling a strong non-cool-core group.
Analyzing the poor and massive cluster simulations (squares, triangles), 
we see no similarity breaking, due to the larger $E_{\rm b}$. The maximum deviation in the core-included relation is highly limited, $\sim\,$0.17/0.08 dex in $L_{\rm x}$ and $\sim\,$0.03/0.015 in $T_{500}$, for the poor and massive cluster, respectively. Excising the core stifles the scatter by at least 1/3.
The powerful AGN heating has again the side effect of destroying the core ($t_{\rm cool}\gg t_{\rm H}$).
Only the massive cluster can partially recover after several Gyr.
Cool cores are common in the universe, hence this type of breaking
should be rare. Notice that combining the two feedback mechanisms aggravates the core overheating.

The global $L_{\rm x}-T_{\rm x}$ property seems overall more likely linked to a primordial imprint that the group/cluster experienced, rather than an internal breaking after formation.
However, we note that when the external `pre-heating' at high redshift  -- whose agency is still unclear -- is high enough to bring $L_{\rm x}$ consistent with observations, the gas entropy becomes usually too high, inhibiting cool cores to form ab initio (e.g., \citealt{Brighenti:2001}).

\section[]{Conclusions}  \label{s:disc}
\noindent
We showed that AGN feedback has severe difficulty in breaking the self-similarity of galaxies, groups, and clusters, in a consistent way.
Via high-resolution 3D simulations, we isolated the impact of the two common regimes of AGN feedback
on the principal scaling relation $L_{\rm x} - T_{\rm x}$.
\begin{itemize}
\item
Self-regulated kinetic feedback prevents the similarity breaking, inducing a limited scatter ($\lta\,$0.1 dex). 
Self-regulation allows to properly quench cooling flows preserving the cool-core structure; avoiding overheating translates thus in a modest central gas evacuation, maintaining low core cooling times ($t_{\rm cool} < 7$ Gyr) and avoiding the $L_{\rm x}-T_{\rm x}$ breaking at any halo scale. Since the majority of observed systems display a cool core (\S\ref{s:intro}), this mode should represent the long-term maintenance phase of AGN feedback.\\

\item
An impulsive quasar thermal blast, injecting the total energy of a typical SMBH, 
is able to break the core-included $L_{\rm x}-T_{\rm x}$ at scales $T_{500}\lta1$ keV (where $E_{\rm AGN}\gta E_{\rm b}$).
However, after full breaking, the system is almost devoid of gas, also at large radii, in contradiction with the core-excised relation.
Even with partial evacuation ($M_{\rm vir}\;$$\gta\;$$5\times10^{13}\ \msun$), the central $t_{\rm cool}$ is raised to several times the Hubble time. In clusters, the scatter is again limited, $\lta0.2$ dex. The imprint of the thermal blast is indelible, morphing the system into a perennial non-cool-core object. If existent, such a mechanism should be rare or occurring at very high redshift.
\end{itemize}

Breaking self-similarity via inside-out heating means to evacuate most of the gas from the region $\lta\,$$R_{500}$. Since central $t_{\rm cool}\propto n_0^{-1}$, lowering the gas density by one order of magnitude
at large radii implies decreasing the core density $n_0$ by at least $10\times$ more, inducing $t_{\rm cool}\gg t_{H}$. The problem is further aggravated by the increase of temperature ($t_{\rm cool}\propto T^{1/2}$).
The direct action of AGN feedback is to lower the luminosity and heating the gas, not moving the system parallel to the $L_{\rm x}-T_{\rm x}$ fit.
Overall, AGN feedback appears naturally suited to regulate the thermodynamical state of cosmic systems, in the core, but not over large radii ($r\gta0.2\,R_{500}$). We remark that any feedback mechanism, that is able to break the self-similarity, needs to properly solve the cooling flow problem.

In a forthcoming work, we discuss other heating models, parameters, and scaling relations.
We found, nevertheless, that AGN feedback models fall in the two archetypal categories presented here: self-regulated heating, preventing the breaking, or strong impulsive heating, which breaks the scaling relations but destroys the core.
For instance, 
using either thermal or kinetic feedback with self-regulation has the same minor impact on the scaling relations (\S\ref{s:cold}).
Injecting energy in the center or at a distance of a few 10 kpc ($\lta0.05\,R_{500}$) has also the same effect, considering that the feedback must affect extremely large regions ($R_{500}$, several 100s kpc); a too distant injection allows instead central runaway cooling. Further, boosting $\epsilon$ transforms the self-regulated feedback into the quasar-like blast; vice versa, diminishing the impulsive $E_{\rm AGN}$ slightly lowers $t_{\rm cool}$, but prevents the similarity breaking. 
In other words, even with a different parametrization of the archetypal AGN 
feedback models, breaking self-similarity implies breaking the cool core.

\section*{Acknowledgments}
\noindent
The FLASH code was in part developed by the DOE NNSA-ASC OASCR Flash center at the University of Chicago. 
MG is grateful for the financial support provided by the Max Planck Fellowship.
SE and FB acknowledge financial contribution from ASI-INAF I/009/10/0, PRIN INAF 2012,
PRIN MIUR 2010LY5N2T.
High-performance computing resources were provided by the NASA/Ames HEC Program 
(SMD-13-3935, SMD-13-4373, SMD-13-4377; Pleiades).
We thank M. Sun for providing the groups data, E. Churazov and the anonymous referee for interesting insights.

\bibliographystyle{apj}

\providecommand{\SortNoop}[1]{}

\end{document}